\documentclass[aip,graphicx,reprint]{revtex4-2}

\usepackage{newtxtext,newtxmath}
\usepackage{hyperref}   
\hypersetup{
    colorlinks=true,
    linkcolor=blue,     
    citecolor=blue,     
    urlcolor=blue       
}

\usepackage{CJKutf8}    
\usepackage{physics}
\usepackage{graphicx}

\begin{document}

\title[]{
  A thermodynamically consistent free-energy lattice Boltzmann model: Incorporating generalized equilibria derived from the color-gradient approach}

\author{Shimpei Saito~(\begin{CJK}{UTF8}{ipxm}齋藤慎平\end{CJK})}
\email[Corresponding author: ]{s.saito@aist.go.jp}
\author{Soumei Baba~(\begin{CJK}{UTF8}{ipxm}馬場宗明\end{CJK})}
\author{Naoki Takada~(\begin{CJK}{UTF8}{ipxm}高田尚樹\end{CJK})}
\affiliation{
  Research Institute for Energy Efficient Technologies, 
  National Institute of Advanced Industrial Science and Technology (AIST), 1-2-1 Namiki, Tsukuba 3058564, Japan
  }

\date{\today}

\begin{abstract}
We extend the chemical-potential-based free-energy lattice Boltzmann (LB) model of Li \textit{et al}. [\href{https://doi.org/10.1103/PhysRevE.103.013304}{Phys. Rev. E \textbf{103}, 013304 (2021)}] by integrating generalized equilibria, originally formulated for the color-gradient LB model using sixth-order Hermite polynomials [Saito \textit{et al.},  \href{https://doi.org/10.1103/PhysRevE.108.065305}{Phys Rev E \textbf{108}, 065305 (2023)}], into a thermodynamically consistent framework.
Our model is formulated on a three-dimensional D3Q27 lattice with a central-moment collision scheme, simplifying implementation and  improving  Galilean invariance.
Numerical tests, including flat-interface equilibrium, stationary and moving droplets in free space, and wetting on solid surfaces, confirm the model's capability to accurately simulate multiphase phenomena while maintaining strict thermodynamic consistency.
\end{abstract}

\maketitle 

\section{Introduction}
The lattice Boltzmann (LB) method is a lattice-based computational fluid dynamics (CFD) approach that models fluid motion through iterative collision and streaming processes of particle distribution functions.  
Compared to traditional CFD methods, the LB method follows a relatively simple time evolution algorithm, making it easier to implement and highly suitable for parallel computing due to its entirely local collision operations~\citep{Chen1998-mm,Kruger2017-ux}.
Furthermore, by properly incorporating external forces, the LB method effectively models multiphase flows and non-ideal fluids, enabling accurate simulations of phenomena involving surface tension and phase separation~\citep{Hosseini2023-nj}.

Several variants of the LB method have been proposed for simulating multiphase flows, such as color-gradient~\citep{Gunstensen1991-wv,Grunau1993-ia}, pseudopotential~\citep{Shan1993-cx,Shan1994-wm}, free-energy~\citep{Swift1995-wb,Swift1996-zh}, and phase-field~\citep{He1999-tm,He1999-pp} models. 
Among these, the free-energy LB model, initially introduced by Swift \textit{et al.}~\citep{Swift1995-wb,Swift1996-zh}, employs thermodynamic considerations through a free-energy framework.
This model modifies the second-order moment of the equilibrium distribution function to include a non-ideal thermodynamic pressure tensor, allowing phase separation governed by a non-ideal equation of state (EOS), such as the van der Waals EOS.
However, the original free-energy model suffered from Galilean invariance issues, subsequently addressed by ~\citet{Pooley2008-rh} through modifications to the equilibrium distribution function, reducing spurious velocities and enhancing physical consistency.
Alternatively, instead of modifying the equilibrium distribution function directly as in Ref.~\citep{Pooley2008-rh}, thermodynamic consistency can also be enhanced by incorporating appropriate forcing terms~\citep{Mazloomi_M2015-ng,Wen2017-om,Wen2020-hn}.

Addressing the issue of thermodynamic inconsistencies arising from numerical errors at discrete lattice levels, Li \textit{et al.}~\citep{Li2021-ar} proposed an improved free-energy LB scheme.
Their approach modifies the EOS, thereby eliminating force discretization errors, achieving better thermodynamic consistency, and significantly reducing spurious currents.
Additionally, Yu \textit{et al.}~\citep{Yu2021-sp} further enhanced the chemical-potential-based free-energy model by introducing thermodynamically consistent boundary conditions for wetting on solid surfaces, thus addressing inconsistencies found in previous boundary schemes~\citep{Wen2017-om}.

The equilibrium distribution function used by Li \textit{et al.}~\citep{Li2021-ar} is based on a third-order Hermite polynomial expansion of the Maxwellian distribution~\citep{Li2012-vr}. 
To correct diagonal elements of third-order moments inadequately adjusted by the equilibrium distribution alone, they introduced additional correction terms through a forcing term. 
Interestingly, the equilibrium distribution function used in the color-gradient LB model has analogous issues, and thus modifications and corrections proposed in Ref.~ \citep{Li2012-vr} are equally beneficial in improving the Galilean invariance of the color-gradient model~\citep{Ba2016-ve,Wen2019-jc}.

Recently, \citet{Saito2023-qd} proposed generalized equilibria derived from sixth-order Hermite polynomials for the color-gradient LB model, significantly improving its Galilean invariance. 
While their equilibrium distribution functions initially appear complex, transforming these functions into central-moment (CM) space significantly simplifies them. 
The CM-based LB method, originally developed by Geier \textit{et al.}~\citep{Geier2006-jt}, performs collisions in the comoving reference frame, removing dependencies on translational motion and reducing Galilean invariance issues~\citep{Coreixas2019-ce}. 
Setting different relaxation rates for various order moments further enhances numerical stability and performance.

In this study, we demonstrate the applicability of generalized equilibria proposed by \citet{Saito2023-qd} for the chemical-potential-based free-energy LB model developed by Li \textit{et al.}~\citep{Li2021-ar}. 
Through comprehensive numerical tests, we validate the model's ability to accurately simulate thermodynamically consistent liquid-vapor phase separation, stationary and moving droplet, and wetting phenomena involving solid surfaces and three-phase contact lines.

\section{Theory and method}

\subsection{Free-energy theory}
The thermodynamic properties of an isothermal single-component liquid-vapor system can be described using the classical second-gradient theory. 
The corresponding Helmholtz free-energy functional is expressed as~\citep{van-der-Waals1979-qs,Rowlinson1982-ip,Swift1995-wb}
\begin{equation}
    \mathcal{F}(\rho,\nabla \rho) = \int{ \qty[ \psi(\rho) + \frac{\kappa}{2} |\nabla \rho|^2 ] }\mathrm{d} \vb{x},
\end{equation}
where $\rho$ denotes the fluid density. 
The first term, $\psi(\rho)$, represents the bulk free-energy density, while the second term $\frac{\kappa}{2}|\nabla \rho|^2$ accounts for the interfacial free energy density arising from non-local molecular interactions. 
Here, $\kappa$ is a positive interfacial free energy coefficient related to surface tension and interfacial thickness.
The first variation of the free energy functional with respect to the density yields the chemical potential $\mu$:
\begin{equation}
    \mu \equiv \frac{\delta \mathcal{F}}{\delta \rho} = \mu_0 - \kappa \nabla ^2 \rho,
\end{equation}
where the bulk chemical potential is defined as $\mu_0 \equiv \partial \psi/\partial \rho$, and the second term represents the capillary contribution involving the Laplacian of the density.

The bulk free energy density $\psi(\rho)$ determines the diagonal component of the pressure tensor, expressed as:  
\begin{equation}
    p = p_0 - \kappa \rho \nabla^2 \rho - \frac{\kappa}{2} |\nabla \rho|^2,
\end{equation}  
with the general expression for the EOS given by
\begin{equation}
\begin{split}
    p_0 & = \rho \psi'(\rho) - \psi(\rho) \\
    &= \rho \mu_0 - \psi(\rho). \label{eq:general_expression_EOS}
\end{split}
\end{equation}
Consequently, the full pressure tensor $\vb{P}$ is defined as
\begin{equation}
\begin{split}
    \vb{P} 
    &= p \vb{I} + \kappa (\nabla \rho \otimes \nabla \rho), \\
    &= \qty( p_0 - \kappa \rho \nabla^2 \rho - \frac{\kappa}{2} |\nabla \rho|^2 )\vb{I} + \kappa (\nabla \rho \otimes \nabla \rho),
\end{split}
\end{equation}  
where $\vb{I}$ denotes the identity tensor.
The excess pressure, representing non-ideal forces relative to an ideal-gas EOS, is expressed explicitly as~\citep{Wen2015-ph}
\begin{equation}
    \vb{F} = -\nabla \cdot \vb{P} + \nabla \cdot \vb{P}_0, \label{eq:nonideal_force_1}
\end{equation}  
where the ideal-gas EOS is $\vb{P}_0 = \rho c_s^2 \vb{I}$, and $c_s$ denotes the speed of sound.
Applying thermodynamic identities leads to the relation $\nabla \cdot \vb{P} = \rho \nabla \mu$, enabling Eq.~(\ref{eq:nonideal_force_1}) to be reformulated into the potential form~\citep{Guo2021-pp}:
\begin{equation}
    \vb{F} = -\rho \nabla \mu + \nabla (\rho c_s^2). \label{eq:nonideal_force_2}
\end{equation} 
This form explicitly highlights the relationship between the thermodynamic force and chemical potential gradients.

\subsection{Chemical-potential-based free-energy LB model}

\begin{figure}[bt]
    \centering
    \includegraphics[width=0.60\linewidth]{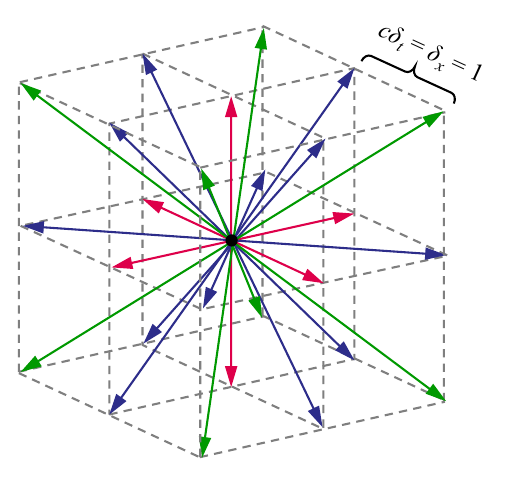}
    \caption{
      Schematic illustration of the D3Q27 lattice used in this study.
      Arrows indicate lattice velocity directions with magnitudes of $|\vb{c}_i|=1$ (red), $|\vb{c}_i|=\sqrt{2}$ (blue), and $|\vb{c}_i|=\sqrt{3}$ (green).
    }
    \label{fig:D3Q27}
\end{figure}

We now introduce the LB equation incorporating a non-ideal external force. 
The LB model employed in this study builds upon the chemical-potential-based free-energy LB model developed by Li \textit{et al.}~\citep{Li2021-ar}.
While their model was originally formulated on a two-dimensional D2Q9 lattice, our approach extends this framework to three dimensions, employing a D3Q27 lattice ($i = 0,1,\dots,26$), as shown in Fig.\ref{fig:D3Q27}.
In the D3Q27 lattice, the speed of sound is defined as $c_s = c/\sqrt{3}$, with $c$ being the lattice velocity~\citep{He1997-ur}. 
Furthermore, whereas Li \textit{et al.} utilized a raw-moment-based (RM) multi-relaxation-time (MRT) collision model, our approach implements a CM-based  MRT collision scheme.

The lattice BGK equation~\citep{Qian1992-wy}, incorporating the effects of external forces, is given by~\citep{Guo2002-qj}:
\begin{equation}
\begin{split}
    f_i(\vb{x} + \vb{c}_i \delta_t, t + \delta_t)  = f_i (\vb{x}, t)  - \frac{1}{\tau} \qty(f_i - f_i^{\mathrm{eq}}) + \delta_t F_i, \label{eq:LBE}
\end{split}
\end{equation}
where $f_i(\vb{x},t)$ is the distribution function of particles traveling with velocity $\vb{c}_i = [c_{ix}, c_{iy}, c_{iz}]^\top$ at position $\vb{x} = [x,y,z]^\top$ and time $t$.
As commonly practiced in LB methods, we set $\delta_t = c = 1$.
Here, $f_i^{\mathrm{eq}}$ denotes the equilibrium distribution function, and $\tau$ is the non-dimensional relaxation time. The term $F_i$ accounts for the external force contribution.
The LB equation~(\ref{eq:LBE}) can be split into two separate steps: the collision step,
\begin{equation}
    f_i^* = f_i (\vb{x}, t)  - \frac{1}{\tau} \qty(f_i - f_i^{\mathrm{eq}}) + \delta_t  F_i, \label{eq:collision_step}
\end{equation}
and the streaming step,
\begin{equation}
    f_i(\vb{x} + \vb{c}_i \delta_t, t + \delta_t)  = f_i^*, \label{eq:streaming_step}
\end{equation}
where $f_i^*$ represents the post-collision distribution functions.
The macroscopic density and velocity are calculated as:
\begin{equation}
    \rho = \sum_i f_i, \quad 
    \rho \vb{u} = \sum_{i}  f_i\vb{c}_i + \frac{\delta t}{2} \vb{F}.
\end{equation}
Following Li \textit{et al.}\citep{Li2021-ar}, the term $\rho c_s^2$ in Eq.(\ref{eq:nonideal_force_2}) is replaced with the modified pressure $p_m = (1 + \mu)/3$. With this modification, the non-ideal force is rewritten as:
\begin{equation}
    \vb{F} =\nabla p_m - \rho \nabla \mu = \qty( \frac{1}{3} - \rho)\nabla \mu. \label{eq:nonideal_force_3}
\end{equation} 

To ensure that the second-order moment relation $\sum_i f_i^\mathrm{eq} c_{i\alpha} c_{i\beta} = p_m \delta_{\alpha \beta}+ \rho u_\alpha u_\beta$ is satisfied, the standard equilibrium distribution function in the LB method must be modified accordingly.
In this study, we adopt the equilibrium distribution function originally derived for the color-gradient LB model by \citet{Saito2023-qd}, defined as:
\begin{equation}
    f_i^{\mathrm{eq}} = g_i^{\mathrm{eq},6} + (p_m - \rho c_s^2) \Phi_i, \label{eq:equilibria_in_general_form}
\end{equation}
where $g_i^\mathrm{eq,6}$ represents the standard equilibrium distribution function expanded to sixth-order $O(u^6)$ using Hermite polynomials~\citep{De_Rosis2019-bq}
\begin{widetext}
\begin{equation}
  \begin{split}
    g_i^{\mathrm{eq},6} = & ~ \rho w_i \left[  1 + 
      \frac{u_x H_{i100} + u_y H_{i010} + u_z H_{i001}}{c_s^2} \right.  \\
    & \left. 
    + \frac{u_x^2 H_{i200} + u_y^2 H_{i020} + u_z^2 H_{i002} 
      + 2( 
          u_x u_y H_{i110} 
        + u_y u_z H_{i011}
        + u_x u_z H_{i101}
      )}{2c_s^4} \right.\\
    & \left.  + \frac{u_x^2 u_y H_{i210} 
      + u_x^2 u_z H_{i201} 
      + u_x u_y^2 H_{i120} 
      + u_x u_z^2 H_{i102} 
      + u_y u_z^2 H_{i012} 
      + u_y^2 u_z H_{i021}  
      + 2u_x u_y u_z H_{i111}}{2c_s^6} \right.\\
      & \left. 
      + \frac{u_x^2 u_y^2 H_{i220}
        + u_x^2 u_z^2 H_{i202}
        + u_y^2 u_z^2 H_{i022}
        + 2(
            u_x u_y u_z^2 H_{i112}
          + u_x u_y^2 u_z H_{i121}
          + u_x^2 u_y u_z H_{i211})}{4c_s^8} \right.\\
    & \left. 
      + \frac{u_x^2 u_y u_z^2 H_{i212} 
        + u_x^2 u_y^2 u_z H_{i221}  
        + u_x u_y^2 u_z^2 H_{i122}}{4c_s^{10}} 
      + \frac{u_x^2 u_y^2 u_z^2  H_{i222}}{8c_s^{12}}  
      \right] \label{eq:equilibria_for_standard_LBM_6th}
  \end{split}
\end{equation}
where $H_{i\alpha\beta\gamma}$ denotes the Hermite polynomial defined in Appendix B of Ref.~\citep{Saito2023-qd}.
The correction operator $\Phi_i$ is expressed as follows:
\begin{equation}
\begin{split}
  \Phi_i = & ~ 
    E_i + w_i \left[ 
    \frac{u_x (H_{i120} + H_{i102}) + u_y (H_{i210} + H_{i012}) + u_z (H_{i201} + H_{i021})}{2c_s^6} \right. \\
     & \left. + \frac{(u_x^2 + u_y^2)  H_{i220} + (u_y^2 + u_z^2)  H_{i022} + (u_x^2 + u_z^2)  H_{i202} + 2(u_y  u_z H_{i211} + u_x u_z H_{i121} + u_x u_y H_{i112})}{4c_s^8}   \right.\\
     & \left.  + \frac{u_x (u_y^2 + u_z^2 -c_s^2) H_{i122} + u_y (u_x^2 + u_z^2 -c_s^2) H_{i212} + u_z (u_x^2 + u_y^2 -c_s^2) H_{i221}}{4c_s^{10}}  \right.\\
     & \left.  + \frac{( u_x^2 u_y^2 + u_y^2 u_z^2 + u_x^2 u_z^2 -c_s^2(u_x^2 + u_y^2 + u_z^2) )H_{i222}}{8c_s^{12}}  \right], 
     \label{eq:New_Equilibriua_in_general_form}
\end{split}
\end{equation}
\end{widetext}
where 
\begin{equation}
\begin{split}
    E_i = w_i 
    & 
    \left(\frac{H_{i200}+H_{i020}+H_{i002}}{2c_s^4} \right. \\
    &  \left. - \frac{H_{i220}+H_{i022}+H_{i202}}{4c_s^6}
    + \frac{H_{i222}}{8c_s^8}
    \right).
    \label{eq:isotropic_operator}
\end{split}
\end{equation}
is the isotropic operator introduced to ensure interface isotropy~\citep{Lafarge2021-ce}.
In Eqs.~(\ref{eq:equilibria_for_standard_LBM_6th})--(\ref{eq:isotropic_operator}), the standard lattice weights for the D3Q27 lattice are given by: 
\begin{equation}
    w_i =
    \begin{cases}
        8/27, & \text{for }  |\vb{c}_i| = 0, \\
        2/27, & \text{for }  |\vb{c}_i| = 1, \\
        1/54, & \text{for }  |\vb{c}_i| = \sqrt{2}, \\
        1/216, & \text{for } |\vb{c}_i| = \sqrt{3}. \label{eq:lattice_weights}
    \end{cases}
\end{equation}  
Although implementing this equilibrium distribution function directly in phase space is cumbersome, its equilibrium CMs exhibit a significantly simpler form, as demonstrated in Ref.~\citep{Saito2023-qd}.
Therefore, in this study, the collision operation is performed directly in the CM space.

The transformation from phase space to CM space is conducted in two steps: first, the distribution functions are converted into RMs, and subsequently, these RMs are mapped onto CMs. 
The definitions of the RMs and CMs are given by~\citep{Geier2006-jt}:
\begin{equation}
    m_{\alpha\beta\gamma} = \sum_i f_i c_{ix}^\alpha c_{iy}^\beta c_{iz}^\gamma,
    \label{eq:raw_moments}
\end{equation}
and 
\begin{equation}
    k_{\alpha\beta \gamma} = \sum_i f_i (c_{ix} - u_x)^\alpha 
    (c_{iy} - u_y)^\beta 
    (c_{iz} - u_z)^\gamma,
    \label{eq:central_moments}
\end{equation}
where the indices $\alpha$, $\beta$, and $\gamma$ each independently take integer values from zero to two, resulting in a total of $3^3=27$ moments.
Although the CMs can be computed directly from Eq.~(\ref{eq:central_moments}), using RMs defined by Eq.~(\ref{eq:raw_moments}) reduces the computational cost. 
Explicit expressions for converting between the distribution functions, RMs, and CMs can be obtained by running the script \texttt{get\_f\_m\_k\_relations.jl}, as provided in the Supplemental Material.

Analogous to Ref.\citep{Saito2023-qd}, the equilibrium CMs corresponding to Eqs.~(\ref{eq:equilibria_for_standard_LBM_6th})--(\ref{eq:isotropic_operator}) are computed explicitly as:
\begin{equation}
\begin{split}
  k_{000}^{\mathrm{eq}} =  &~ \rho , \\
  k_{200}^{\mathrm{eq}} = k_{020}^{\mathrm{eq}} = k_{002}^{\mathrm{eq}} = &~ p_m   ,  \\
  k_{220}^{\mathrm{eq}} = k_{202}^{\mathrm{eq}} = k_{022}^{\mathrm{eq}} =  &~ p_mc_s^2 , \\
  k_{222}^{\mathrm{eq}} = &~  p_m c_s^4 .
\end{split}
\end{equation}
All equilibrium CMs, except these, are identically zero, greatly simplifying the implementation.
By adopting the CM-based approach, the complicated direct implementation of Eqs.~(\ref{eq:equilibria_for_standard_LBM_6th})--(\ref{eq:isotropic_operator}) is no longer required.

As a result, the collision step is formulated as follows~\citep{Saito2023-qd}:
\noindent \textbf{First order:}
\begin{equation}
\begin{split}
  k_{100}^{*} &= (1 - \omega_0)k_{100} + \qty(1 - \frac{\omega_0}{2})F_x ,  \\
  k_{010}^{*} &= (1 - \omega_0)k_{010} + \qty(1 - \frac{\omega_0}{2})F_y ,  \\
  k_{001}^{*} &= (1 - \omega_0)k_{001} + \qty(1 - \frac{\omega_0}{2})F_z , 
  \label{eq:collision_step_in_general_form_1st}
\end{split}
\end{equation}
\begin{widetext}
\noindent \textbf{Second order:}
\begin{equation}
\begin{split}
  k_{110}^{*} &= (1 - \omega_1)k_{110},  \\
  k_{011}^{*} &= (1 - \omega_1)k_{011},  \\
  k_{101}^{*} &= (1 - \omega_1)k_{101},  \\
  k_{200}^{*} - k_{020}^{*} &= (1 - \omega_1)(k_{200} - k_{020}) + \qty(1 - \frac{\omega_1}{2})(Q_{200} - Q_{020}),  \\
  k_{200}^{*} - k_{002}^{*} &= (1 - \omega_1)(k_{200} - k_{002}) + \qty(1 - \frac{\omega_1}{2})(Q_{200} - Q_{002}),  \\
  k_{200}^{*} + k_{020}^{*} + k_{002}^{*} &= (1 - \omega_2)(k_{200} + k_{020} + k_{002}) + 3 \omega_2 p_m + \qty(1 - \frac{\omega_2}{2})(Q_{200} + Q_{020} + Q_{002}),  \\
  \label{eq:collision_step_in_general_form_2nd}
\end{split}
\end{equation}
\end{widetext}
\noindent \textbf{Third order:}
\begin{equation}
\begin{split}
  k_{120}^{*} + k_{102}^{*} &= (1-\omega_3)(k_{120} + k_{102}) 
  + 2\qty(1-\frac{\omega_3}{2})F_x c_s^2, \\
  k_{210}^{*} + k_{012}^{*} &= (1-\omega_3)(k_{210} + k_{012}) 
  + 2\qty(1-\frac{\omega_3}{2})F_y c_s^2, \\
  k_{201}^{*} + k_{021}^{*} &= (1-\omega_3)(k_{201} + k_{021}) 
  + 2\qty(1-\frac{\omega_3}{2})F_z c_s^2, \\
  k_{120}^{*} - k_{102}^{*} &= (1-\omega_4)(k_{120} - k_{102}), \\
  k_{210}^{*} - k_{012}^{*} &= (1-\omega_4)(k_{210} - k_{012}), \\
  k_{201}^{*} - k_{021}^{*} &= (1-\omega_4)(k_{201} - k_{021}), \\
  k_{111}^{*} &= (1-\omega_5) k_{111}, \\
  \label{eq:collision_step_in_general_form_3rd}
\end{split}
\end{equation}
\begin{widetext}
\noindent \textbf{Fourth order:}
\begin{equation}
\begin{split}
  k_{220}^{*} - 2k_{202}^{*} + k_{022}^{*} &= (1-\omega_6)(k_{220} - 2k_{202} + k_{022})  + \qty(1 - \frac{\omega_6}{2})(Q_{220} -2 Q_{202} + Q_{022}),  \\
  k_{220}^{*} + k_{202}^{*} - 2k_{022}^{*} &= (1-\omega_6)(k_{220} + k_{202} - 2k_{022}) + \qty(1 - \frac{\omega_6}{2})(Q_{220} + Q_{202} -2 Q_{022}),  \\
  k_{220}^{*} + k_{202}^{*} + k_{022}^{*} &= (1-\omega_7)(k_{220} + k_{202} + k_{022}) + 3\omega_7 p_m c_s^2  + \qty(1 - \frac{\omega_7}{2})(Q_{220} + Q_{202} + Q_{022}),  \\
  k_{211}^{*} &= (1-\omega_8)k_{211}, \\
  k_{121}^{*} &= (1-\omega_8)k_{121}, \\
  k_{112}^{*} &= (1-\omega_8)k_{112}, \\
  \label{eq:collision_step_in_general_form_4th}
\end{split}
\end{equation}
\end{widetext}
\noindent \textbf{Fifth order:}
\begin{equation}
\begin{split}
  k_{122}^{*} &= (1-\omega_9)k_{122} + \qty(1 - \frac{\omega_9}{2})F_x c_s^4, \\
  k_{212}^{*} &= (1-\omega_9)k_{212} + \qty(1 - \frac{\omega_9}{2})F_y c_s^4, \\
  k_{221}^{*} &= (1-\omega_9)k_{221} + \qty(1 - \frac{\omega_9}{2})F_z c_s^4, \\
  \label{eq:collision_step_in_general_form_5th}
\end{split}
\end{equation}
\noindent \textbf{Sixth order:}
\begin{equation}
\begin{split}
  k_{222}^{*} &= (1-\omega_{10})k_{222} + \omega_{10} p_m c_s^4.
  \label{eq:collision_step_in_general_form_6th}
\end{split}
\end{equation}
Here, $k_{\alpha\beta\gamma}^*$ and $k_{\alpha\beta\gamma}^{\mathrm{eq}}$ denote the post-collision and equilibrium CMs, respectively.
The relaxation rates $\omega_0, \omega_1, \dots, \omega_{10}$ control the relaxation process.
Among them, $\omega_1$ is related to the kinematic viscosity $\nu$ as $1/\omega_1 = \tau = \rho \nu/( p_m \delta_t) + 1/2$ as in Ref.~\citep{Li2021-ar}, while $\omega_2$ is associated with the bulk viscosity.
The remaining relaxation rates are free parameters and can be chosen within the range $[0,2]$.
In this study, all relaxation rates except for $\omega_1$ are set to unity, further simplifying the right-hand side of Eqs.~(\ref{eq:collision_step_in_general_form_1st})--(\ref{eq:collision_step_in_general_form_6th}).

Despite modifications to the equilibrium distribution function, the diagonal elements of the third-order moments retain the term $\rho c_s^2$ due to the low symmetry of standard lattices~\citep{Saito2023-qd,Wen2019-jc,Li2021-ar}.  
To address this issue, additional correction terms are introduced in Eq.~(\ref{eq:collision_step_in_general_form_2nd}) as external forces acting on the relevant second-order diagonal moments:
\begin{equation}
\begin{split}
    Q_{200} &= -3 \partial_x \qty[(p_m - \rho c_s^2) u_x],\\
    Q_{020} &= -3 \partial_y \qty[(p_m - \rho c_s^2) u_y], \\
    Q_{002} &= -3 \partial_z \qty[(p_m - \rho c_s^2) u_z].
\end{split}
\end{equation}
In addition to these second-order corrections, the following fourth-order correction is applied approximately in Eq.~(\ref{eq:collision_step_in_general_form_4th}):
\begin{equation}
\begin{split}
    Q_{220} &= (Q_{200} + Q_{020})c_s^2,\\
    Q_{202} &= (Q_{200} + Q_{002})c_s^2, \\
    Q_{022} &= (Q_{020} + Q_{002})c_s^2. \label{eq:fourth_order_correction}
\end{split}
\end{equation}
We observed that this treatment improves numerical stability in problems involving curved interfaces compared to flat ones.  
The theoretical validity of this correction should be investigated in a separate study.  
The impact of the correction in Eq.~(\ref{eq:fourth_order_correction}) is analyzed in Sec.~\ref{sec:spherical_droplet}.

After completing the collision step in Eqs.~(\ref{eq:collision_step_in_general_form_1st})--(\ref{eq:collision_step_in_general_form_6th}),  
the CMs are transformed into raw moments and then into the distribution function,  
followed by the streaming step described in Eq.~(\ref{eq:streaming_step}).  
The script provided in the Supplemental Material (\texttt{get\_f\_m\_k\_relations.jl}) facilitates the conversion of moments back into the distribution function.

\section{Numerical tests}

As a fundamental non-ideal fluid model, we adopt the van der Waals (vdW) EOS~\citep{Rowlinson1982-ip}:
\begin{equation}
    p_0 = \frac{\rho R T}{1 - b \rho} - a \rho^2, \label{eq:van_der_Waals}
\end{equation}
where $a$ represents the attractive molecular interaction parameter, $b$ accounts for the volume correction due to finite molecular size, and $R$ is the gas constant.  
By solving the equations $\frac{\partial p_0}{\partial \rho} (\rho_c, T_c) = 0$ and $\frac{\partial^2 p_0}{\partial \rho^2} (\rho_c, T_c) = 0$ at the critical point, the critical density $\rho_c$, temperature $T_c$, and pressure $p_c$ of Eq.~(\ref{eq:van_der_Waals}) are determined as
\begin{equation}
    \rho_c = \frac{1}{3b},~ T_c = \frac{8a}{27Rb},~ p_c = \frac{a}{27b^2}. \label{eq:critical_parameters}
\end{equation}
With the vdW EOS, the corresponding free-energy density and chemical potential are given by~\citep{Wen2017-om}
\begin{equation}
    \psi(\rho) = \rho R T \ln \qty( \frac{\rho}{1 - b\rho}) - a \rho^2,
\end{equation}
\begin{equation}
    \mu = RT \qty[ \ln \qty( \frac{\rho}{1 - b \rho}) + \frac{1}{1 - b\rho} ] - 2a \rho - \kappa \nabla^2 \rho.
\end{equation}

In the following simulations, unless stated otherwise, the parameters are set as  
$a = 9/392$, $b = 2/21$, $R = 1$, and $k = 0.02$.  
From Eq.~(\ref{eq:critical_parameters}), the critical parameters are determined as  
$\rho_c = 7/2$, $T_c = 1/14$, and $p_c = 3/32$.  
The system temperature and kinematic viscosity are set to $T = 0.7 T_c$ and $\nu = 0.15$, except in Sec.~\ref{sec:flat_interface}.  
Under this temperature, the liquid and vapor densities predicted by Maxwell's equal-area rule are  
$\rho_l \approx 7.49$ and $\rho_v \approx 0.448$, yielding a density ratio of $\rho_l/\rho_v \approx 16.7$.  
All results are presented in lattice units.

All simulations were conducted in a three-dimensional computational domain.  
The computational code was implemented in \texttt{CUDA C++}, and the simulations were performed on a single GPU (NVIDIA RTX A4000) with double precision.

\subsection{Flat interface\label{sec:flat_interface}}

\begin{figure}[tb]
    \centering
    \includegraphics[width=0.9\linewidth]{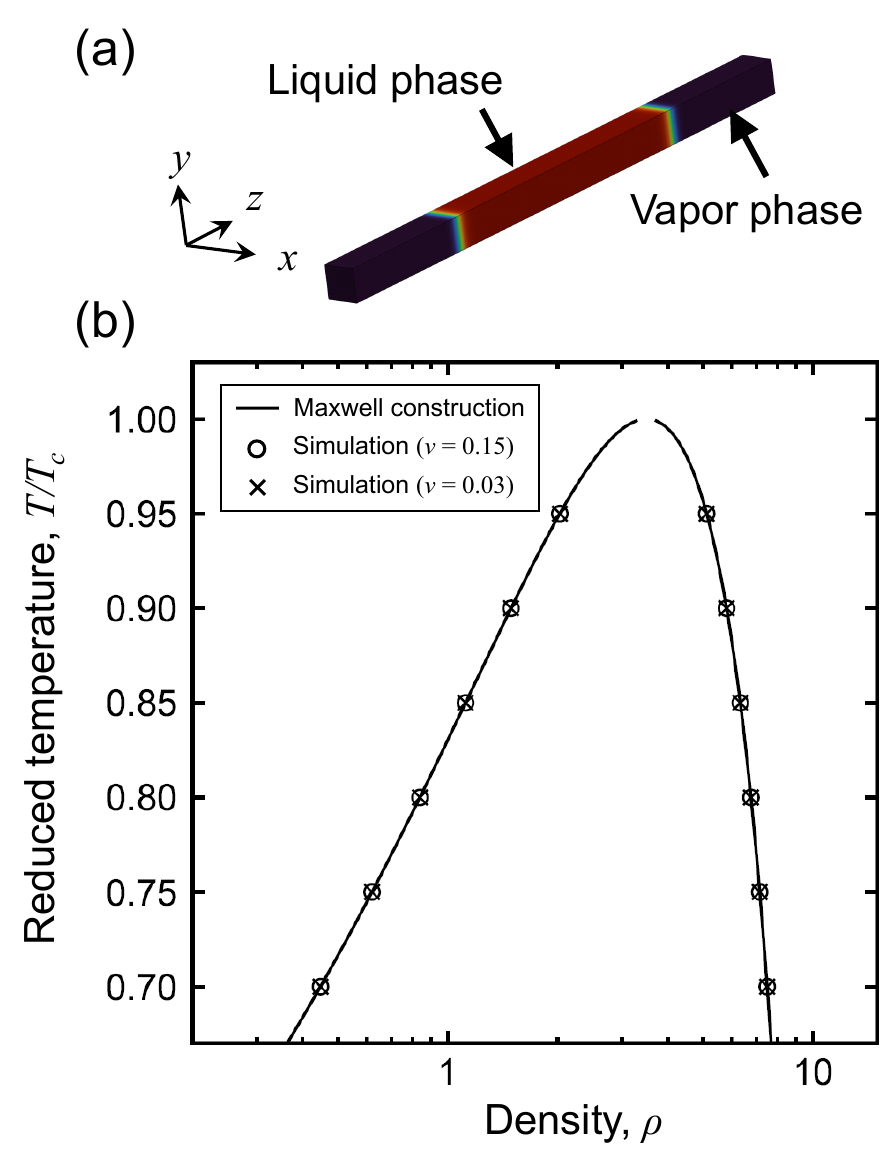}
    \caption{Liquid-vapor equilibrium densities from flat interface simulations (symbols) compared with the coexistence curve from Maxwell's equal-area construction (lines).}
    \label{fig:flat_interface}
\end{figure}

To evaluate the thermodynamic consistency of the present model at equilibrium,  
simulations were conducted for a system with a flat interface.

The computational domain was discretized into a grid of size  
$N_x \times N_y \times N_z = 4 \times 4 \times 128$.  
Initially, flat interfaces were placed at $z = 0.25N_z$ and $z = 0.75N_x$,  
with the central region filled with the liquid phase and the remaining space occupied by the gas phase.  
Periodic boundary conditions were applied in all directions.  
The liquid and vapor phase densities were measured at given reduced temperatures.

Figure~\ref{fig:flat_interface} shows the computational setup and the coexistence curve  
from the simulation results for $\nu = 0.15$ and $\nu = 0.03$.  
The solid line in Fig.~\ref{fig:flat_interface}(b) represents the theoretical curve  
predicted by Maxwell's equal-area construction.  
The simulation results show that the liquid-vapor coexistence curve agrees well  
with the Maxwell construction, regardless of kinematic viscosity.  
The pseudopotential model typically requires parameter tuning to approximately  
achieve thermodynamic consistency~\citep{Li2012-rn}.  
In contrast, the present model naturally satisfies thermodynamic consistency without such tuning.  

\begin{figure}[tb]
  \centering
  \includegraphics[width=1\linewidth]{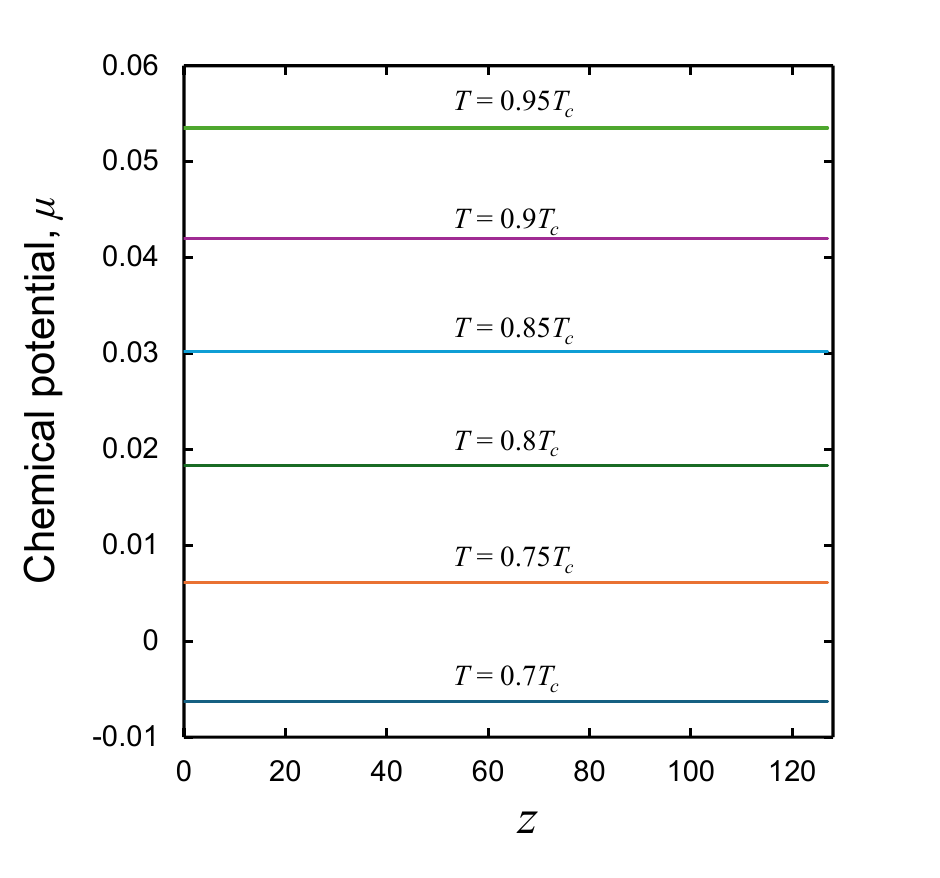}
  \caption{
    Chemical potential distribution along the $z$-axis at different temperatures.
    }
  \label{fig:flat_interface_chemical_potential}
\end{figure}

Additionally, thermodynamic consistency can be assessed not only  
from the equilibrium densities of the vapor and liquid phases  
but also from the spatial distribution of the chemical potential.  
Figure~\ref{fig:flat_interface_chemical_potential} shows  
the chemical potential distribution along the $z$-axis at equilibrium for each temperature condition.  
At all temperatures, the chemical potential remains spatially uniform,  
as required for thermodynamic equilibrium.  
This result further confirms that the present model maintains thermodynamic consistency.

The surface tension is determined from the equilibrium density profile using~\citep{Rowlinson1982-ip,Zhang2000-qs,Kikkinides2008-nj}
\begin{equation}
    \sigma = \kappa \int \qty( \frac{\partial \rho}{\partial z} )^2 \mathrm{d}z.
\end{equation}  
In this study, integration over the range $0$ to $N_z/2$ is sufficient.  
At $T = 0.7T_c$, the surface tension is evaluated as $\sigma = 0.204$.  

\subsection{Droplet in free space\label{sec:spherical_droplet}}

This test examines the role of the fourth-order correction term in Eq.~(\ref{eq:fourth_order_correction})  
in a three-dimensional system with a curved interface.  
It also verifies that the equilibrium density of a droplet with a spherical interface follows the Young--Laplace law.  

The computational domain is discretized into a grid of size $N_x \times N_y \times N_z = 128 \times 128 \times 128$,  
with periodic boundary conditions applied in all directions.  
A droplet of radius $r$ is placed at the center of the domain, with the surrounding region filled with vapor.  
The density field is smoothly initialized as  
\begin{equation}
    \rho = \frac{\rho_l + \rho_v}{2} - \frac{\rho_l - \rho_v}{2} \tanh \qty[ \frac{2(r_0 - r)}{W} ],
\end{equation}
where $W = 5$ represents the initial interface thickness, and $r_0$ is given by  
\begin{equation}
    r_0 = \sqrt{(x-x_0)^2 + (y-y_0)^2 + (z-z_0)^2},
\end{equation}
with $x_0$, $y_0$, and $z_0$ denoting the center of the domain.

\begin{figure*}[tb]
  \centering
  \includegraphics[width=1\linewidth]{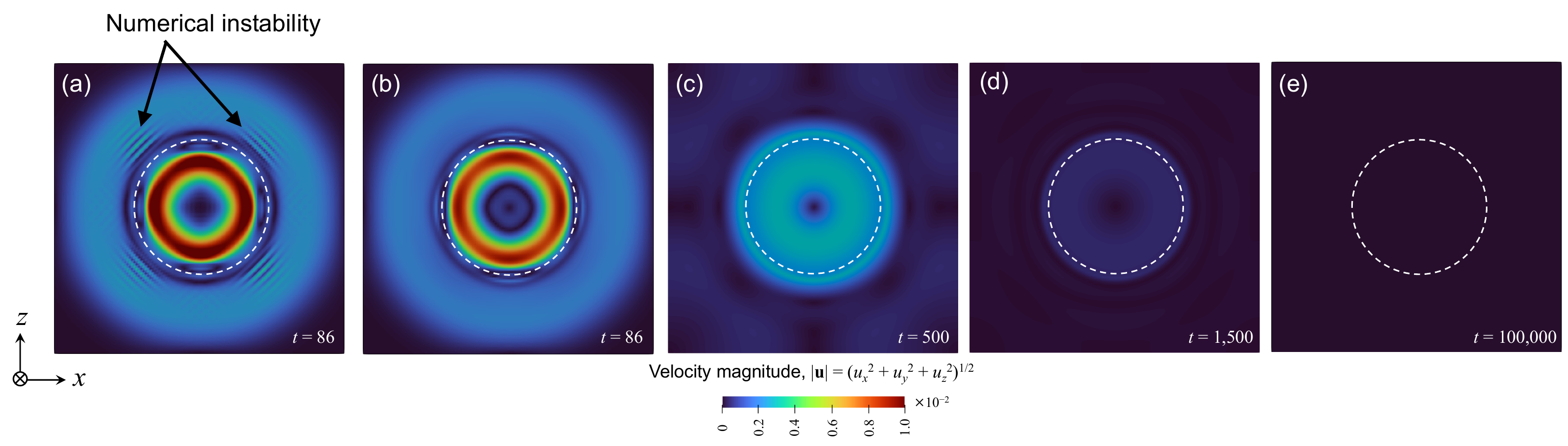}
  \caption{
  Velocity field of a stationary droplet in the $x$-$z$ cross-section:  
  (a) without the fourth-order correction ($t=86$),
  (b) with the correction ($t=86$),
  (c-e) velocity fields with the correction at $t=500$, $1,\!500$, and $100,\!000$, respectively.
  The same velocity scale is used in all panels to illustrate the dissipation process.
  The white dashed line represents the interface position at $(\rho_l + \rho_v)/2$. 
}
  \label{fig:effect_of_correction}
\end{figure*}

A stationary droplet simulation was conducted with $r=30$.  
Figure~\ref{fig:effect_of_correction} illustrates the velocity distribution in the $x$-$z$ cross-section,  
comparing cases with and without the fourth-order correction [Eq.~(\ref{eq:fourth_order_correction})].  
Panels (a) and (b) correspond to the same time step ($t=86$), where (a) excludes the fourth-order correction, while (b) includes it. 
Comparing Figs.~\ref{fig:effect_of_correction}(a) and (b),  
a flow induced by the initial conditions is observed in both cases.  
However, without the fourth-order correction [Fig.~\ref{fig:effect_of_correction}(a)],  
numerical instability developed in the vapor region near the droplet,  
causing the computation to diverge shortly after this time step.
Conversely, when the fourth-order correction is applied,  
the velocity induced by the initial conditions diminishes over time.  
This dissipation process is visualized in Figs.~\ref{fig:effect_of_correction}(c)--(e), which show the velocity field at $t = 500$, $1,\!500$, and $100,\!000$, respectively.
Under these conditions, the maximum spurious velocity remains small,  
reaching approximately $|\vb{u}|_\mathrm{max} = 2.4 \times 10^{-11}$ at time step $t = 100,\!000$.  

\begin{figure}
    \centering
    \includegraphics[width=1\linewidth]{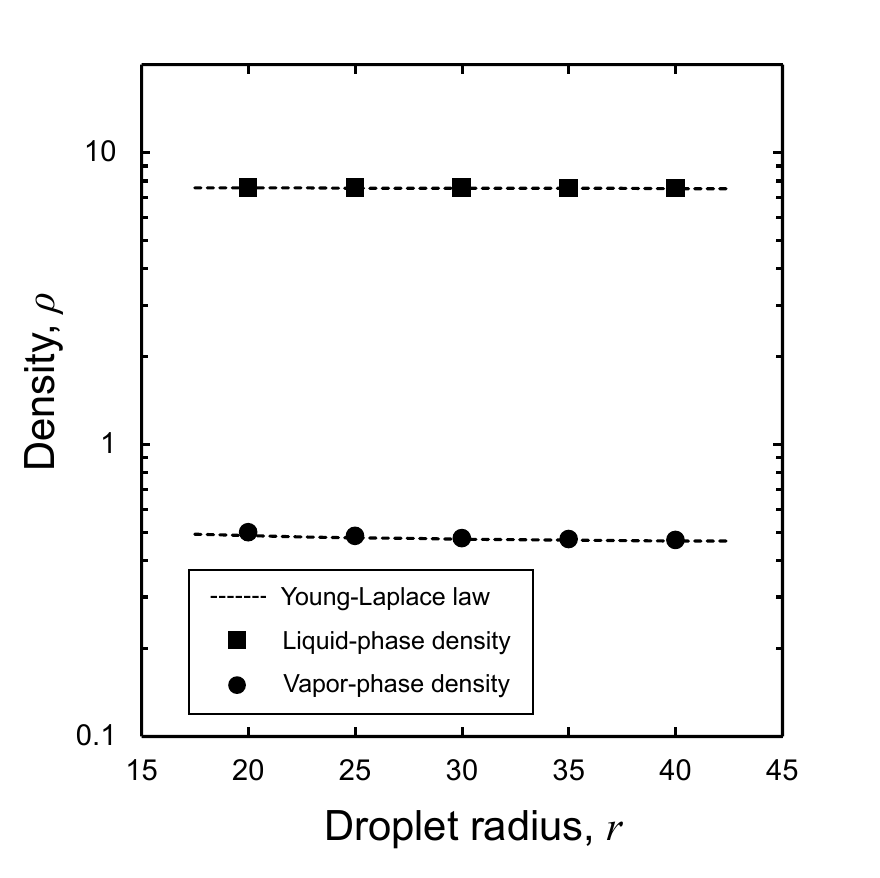}
    \caption{Comparison of liquid and vapor densities from the stationary droplet simulation with the Young--Laplace law~\citep{Jamet2001-kb}.}
    \label{fig:young-laplace}
\end{figure}

In this equilibrium state, the densities inside and outside droplets of different radii  
were measured and compared with the Young--Laplace law,  
using the same method as in Refs.~\citep{Jamet2001-kb,Kikkinides2008-nj,Li2021-ar}.  
As shown in Fig.~\ref{fig:young-laplace}, the simulation results closely match the Young--Laplace law,  
confirming that the surface tension predicted by this model is accurate.

Notably, in the flat interface simulation, the computation remained stable even without the fourth-order correction in Eq.~(\ref{eq:fourth_order_correction}).  
This suggests that the fourth-order correction is crucial for systems with curved interfaces.  
In all subsequent simulations, this correction is applied.

As previously reported~\citep{Inamuro2000-ol,Kalarakis2002-vm},  
the lack of Galilean invariance leads to significant deformation of an initially circular droplet in moving wall simulations.  
To verify the Galilean invariance of the present model, an additional droplet simulation was conducted  
with a slight modification to the boundary conditions of the stationary droplet case.  
Specifically, the top and bottom boundaries were set as moving walls~\citep{Zong2021-ne}  
with a constant velocity of $U = 0.1$, inducing flow inside the domain and causing the droplet to move.

\begin{figure*}[tb]
    \centering
    \includegraphics[width=0.9\linewidth]{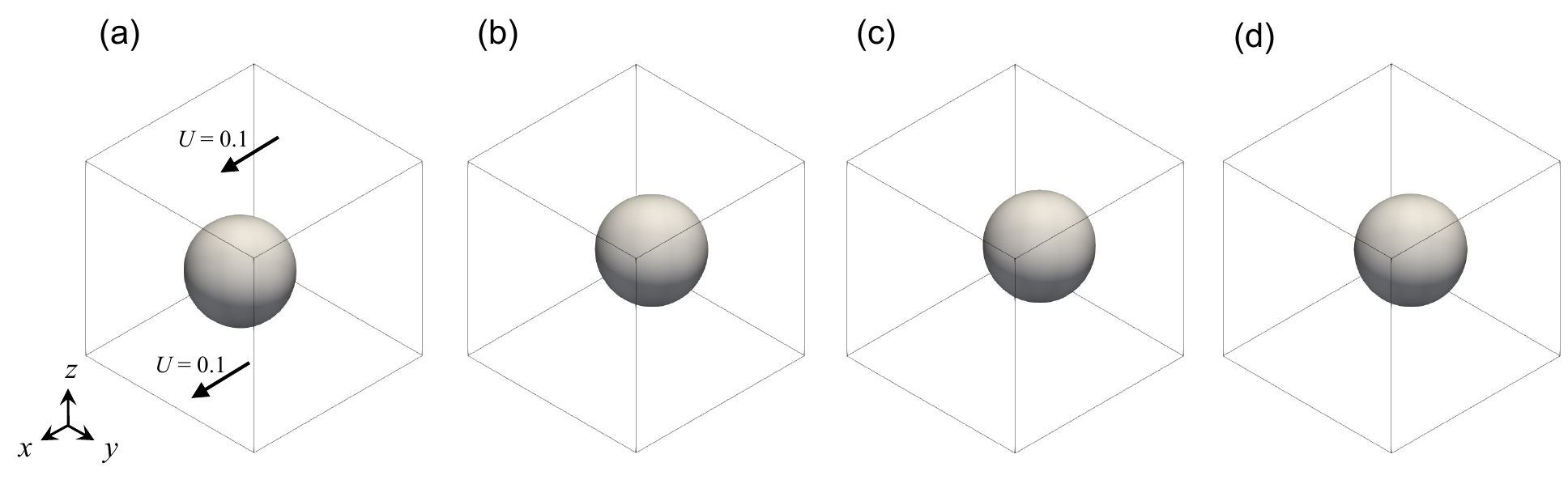}
    \caption{
    Simulation of a moving droplet with top and bottom boundaries set as moving walls at  
    (a) $t = 25,000$, (b) $t = 50,000$, (c) $t = 75,000$, and (d) $t = 100,000$.  
    The interface is defined as $\rho = (\rho_l + \rho_v)/2$.
}
    \label{fig:moving_droplet}
\end{figure*}

Figure~\ref{fig:moving_droplet} shows the droplet interface at each time step up to $t=100,000$.  
The droplet retains its nearly circular shape over time, without undergoing unphysical deformation.  
This confirms an improvement in Galilean invariance of the present chemical-potential-based free-energy model.

\subsection{Droplet on a solid surface}

The numerical tests so far have focused on droplet behavior in free space.  
As a final case incorporating solid boundaries, a droplet wetting simulation is performed.

\begin{figure*}
    \centering
    \includegraphics[width=0.9\linewidth]{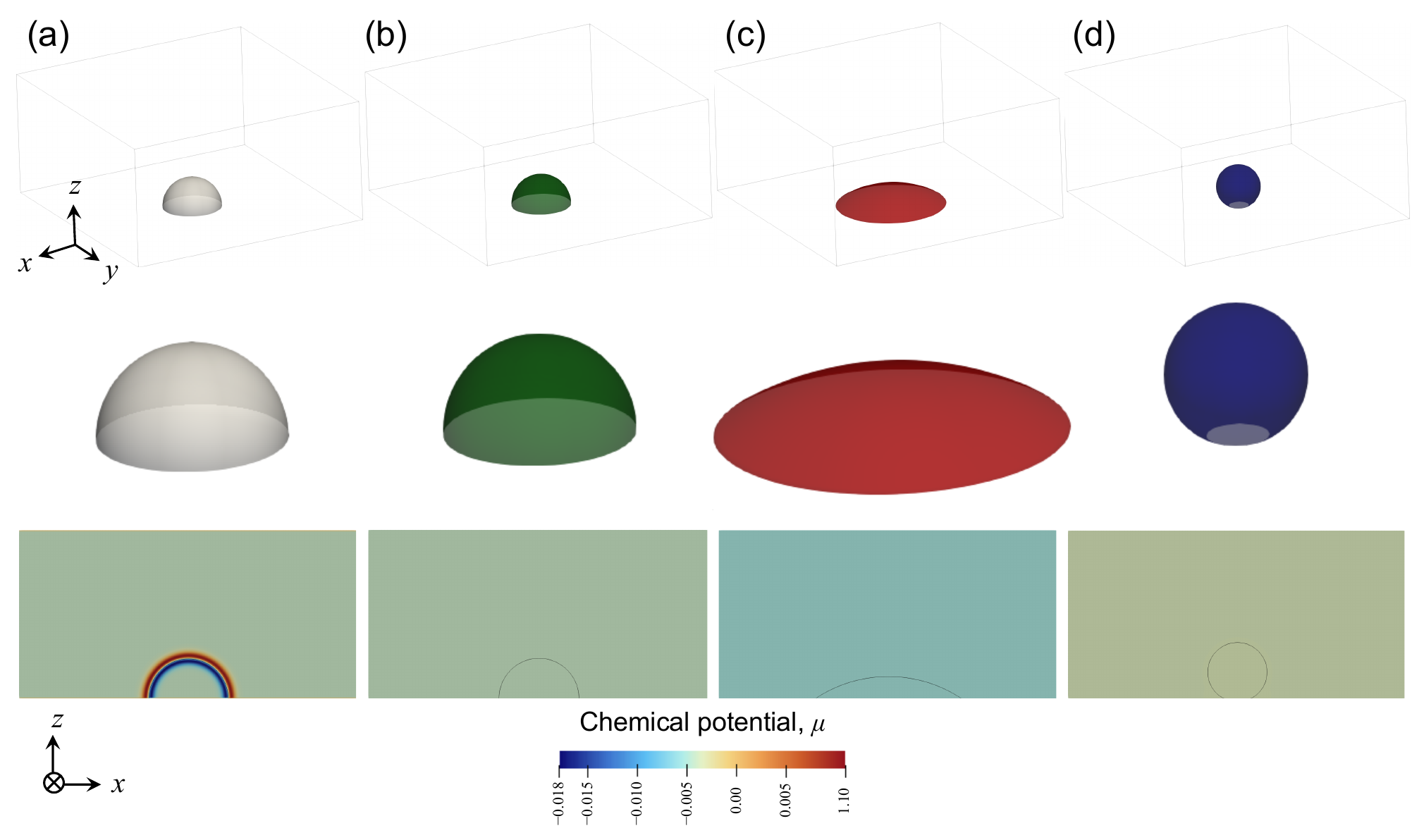}
    \caption{
    Droplet wetting on a solid surface with the improved wetting boundary condition~\citep{Yu2021-sp}:  
    (a) Initial condition,  
    (b) Neutral wetting [$\rho(\vb{x}_s) = \rho_\mathrm{ave}$, $\varphi=1.0$, $\theta = 89.9^\circ$],  
    (c) Hydrophilic state [$\varphi = 1.6$ in Eq.~(\ref{eq:three_phase_contact_line}), $\theta = 33.5^\circ$],  
    (d) Hydrophobic state [$\Delta \rho=1.8$ in Eq.~(\ref{eq:three_phase_contact_line}), $\theta = 153.4^\circ$].  
    The interface is defined as $\rho = (\rho_l + \rho_v)/2$.  
    The bottom row shows the spatial distribution of the chemical potential in the $x$-$z$ cross-section.  
}
    \label{fig:wetting}
\end{figure*}

The computational domain is discretized into a grid of size $N_x \times N_y \times N_z = 128 \times 256 \times 256$.  
A droplet of radius $r = 30$ is placed at the center of the $x$-$y$ plane at $z = 0$,  
with the surrounding region filled with vapor, as shown in Fig.~\ref{fig:wetting}(a).  
The halfway bounce-back scheme~\citep{Kruger2017-ux} is applied at the top and bottom boundaries,  
while periodic boundary conditions are imposed on the remaining boundaries.

The improved wetting boundary scheme proposed by Yu \textit{et al.}~\citep{Yu2021-sp}  
was extended to the D3Q27 lattice and applied to solid surfaces.  
In this approach, the surface chemical potential is defined as  
\begin{equation}
    \mu(\vb{x}_s) = \frac{\sum_i w_i \mu(\vb{x}_s + \vb{c}_i \delta_t) s_w(\vb{x}_s + \vb{c}_i \delta_t)}{\sum_i w_i s_w(\vb{x}_s + \vb{c}_i \delta_t) },
    \label{eq:surface_chemical_potential}
\end{equation}
where $s_w$ is a switch function that takes a value of 1 at fluid nodes and 0 at solid nodes,  
and $w_i$ denotes the lattice weights given in Eq.~(\ref{eq:lattice_weights}).
Note that in the D3Q27 lattice, nine distribution functions pointing toward the interior of the computational domain are unknown.  
Similarly, the average density at each solid node is given by  
\begin{equation}
    \rho_{\mathrm{ave}}(\vb{x}_s) = \frac{\sum_i w_i \rho(\vb{x}_s + \vb{c}_i \delta_t) s_w(\vb{x}_s + \vb{c}_i \delta_t)}{\sum_i w_i s_w(\vb{x}_s + \vb{c}_i \delta_t) },
    \label{eq:surface_density}
\end{equation}
Following Ref.~\citep{Yu2021-sp}, the three-phase contact line region is identified using the criterion:
$
    0.05 \rho_l + 0.95 \rho_v \leq \rho_{\mathrm{ave}}(\vb{x}_s) \leq 0.95 \rho_l + 0.05 \rho_v.
$
Within this region, the surface density is modified as  
\begin{equation}
    \rho(\vb{x}_s) =
    \begin{cases}
        \varphi \rho_{\text{ave}}(\vb{x}_s), & \text{for } \theta \leq 90^\circ, \\
        \rho_{\text{ave}}(\vb{x}_s) - \Delta \rho, & \text{for } \theta > 90^\circ,
    \end{cases}
    \label{eq:three_phase_contact_line}
\end{equation}  
where $\varphi > 1$ and $\Delta \rho > 0$ are constants used to adjust the contact angle.  
For solid nodes other than those in the three-phase contact line region, the density is simply set to $\rho(\vb{x}_s) = \rho_{\text{ave}}(\vb{x}_s)$.  
As a limiter, the surface density given by Eq.~(\ref{eq:three_phase_contact_line}) is constrained within  
$
    0.05 \rho_l + 0.95 \rho_v \leq \rho(\vb{x}_s) \leq 0.95 \rho_l + 0.05 \rho_v.
$

Three types of droplet simulations were conducted using the solid surface treatments in Eqs.~(\ref{eq:surface_chemical_potential})--(\ref{eq:three_phase_contact_line}):  
\begin{itemize}
  \item Neutral wetting, simulated without Eq.~(\ref{eq:three_phase_contact_line}).  
  \item Hydrophilic droplets, simulated by setting $\varphi > 1$.
  \item Hydrophobic droplets, simulated by setting $\Delta \rho > 0$.
\end{itemize}
The simulation results for droplet wetting with the improved wetting boundary scheme are summarized in Fig.~\ref{fig:wetting}.  
The first and second rows show the droplet interface, with the second row providing a magnified view.  
The results indicate that, without Eq.~(\ref{eq:three_phase_contact_line}) [Fig.~\ref{fig:wetting}(b)],  
the contact angle is $89.9^\circ$, corresponding to nearly neutral wetting,  
which is not significantly different from the initial condition [Fig.~\ref{fig:wetting}(a)] in shape.

When $\varphi = 1.6$, the droplet spread over the solid surface, exhibiting a hydrophilic state  
with a contact angle of $33.5^\circ$ [Fig.~\ref{fig:wetting}(c)].  
Conversely, when $\Delta \rho = 1.8$, the droplet retained a nearly spherical shape on the solid surface,  
exhibiting a hydrophobic state with a contact angle of $153.4^\circ$ [Fig.~\ref{fig:wetting}(d)].  
These results demonstrate that both hydrophilic and hydrophobic behaviors can be accurately captured  
using the surface density and surface chemical potential treatments in  
Eqs.~(\ref{eq:surface_chemical_potential})--(\ref{eq:three_phase_contact_line}).

Finally, the spatial distribution of the chemical potential is examined, shown in the bottom row of Fig.~\ref{fig:wetting}.  
At the initial stage of the computation [Fig.~\ref{fig:wetting}(a)], an inhomogeneous chemical potential is observed near the interface.  
In contrast, at equilibrium [Figs.~\ref{fig:wetting}(b)–(d)], the chemical potential becomes spatially uniform, regardless of the contact angle.  
This result is consistent with thermodynamic equilibrium, where the chemical potential should remain uniform.  
Thus, this chemical-potential-based free-energy model ensures thermodynamically consistent wetting simulations, even in the presence of solid surfaces.

\section{Conclusions}
We developed a chemical-potential-based free-energy LB model using generalized equilibria derived from higher-order Hermite polynomials,
building on the mathematical similarities between the equilibrium distribution functions in the free-energy and color-gradient models.
By employing a CM collision framework, we significantly reduced the complexity of the equilibrium distribution functions while ensuring thermodynamic consistency.

Our numerical validations demonstrated the following key findings:
\begin{itemize}
    \item Accurate reproduction of equilibrium liquid-vapor densities predicted by Maxwell's construction, validating the thermodynamic consistency of the model.
    \item Agreement with the Young--Laplace law for curved interfaces, confirming the reliability of surface tension calculations.
    \item Improved Galilean invariance, as verified by the moving droplet test.
    \item Improved wetting boundary conditions that allow for both hydrophilic and hydrophobic behaviors while maintaining a uniform chemical potential.
\end{itemize}
These findings highlight the importance of the fourth-order correction in ensuring numerical stability and accuracy,  
particularly for curved interfaces in three dimensions.
In particular, the correction plays a crucial role in maintaining numerical stability near curved liquid-vapor interfaces.
Furthermore, the proposed model serves as a reliable and thermodynamically consistent approach for simulating multiphase flows.  
The incorporation of generalized equilibria and appropriate correction terms into the free-energy framework ensures numerical consistency,  
which is essential for accurately capturing interface dynamics.

Despite these advantages, some challenges remain.  
For example, the current model assumes a fixed EOS, which may limit its applicability to more complex thermodynamic systems.  
Future work could explore extensions to multi-component systems, as well as the incorporation of dynamic wetting effects under non-equilibrium conditions.

Overall, this study establishes a solid foundation for further advancements in thermodynamically consistent LB models for multiphase flows.


\section*{Author Declarations}
\subsection*{Author Contributions}
\textbf{Shimpei Saito:} Software; Data curation; Validation; Conceptualization; Writing -- original draft;  Project administration; Funding acquisition.
\textbf{Soumei Baba:} Validation; Writing -- review \& editing.
\textbf{Naoki Takada:} Methodology; Writing -- review \& editing.

\section*{Data availability}
Data available on request from the authors.

\begin{acknowledgments}
  This work was supported by the New Energy and Industrial Technology Development Organization (NEDO) JPNP14004 and JSPS KAKENHI Grants No. JP22K14201.
  Part of this work was supported by JSPS KAKENHI Grants No. JP20K04297.
\end{acknowledgments}


%

\end{document}